**Quantum Teleportation toward the Quantum Internet: A Concise Review**

*Yang-Bin Ma*[1,2], *Yun-Ru Fan*[1,2,*], *Ri-Yao Song*[1,2], *Ya-Zhou Zhao*[1,2], *Yi-Ye Liu*[3], *Si Shen*[4], *Zi-Hao Zhan*[5], *Yan-Yu Wei*[1], *Kai Guo*[5,*], *Guang-Can Guo*[1,2,6,7], *and Qiang Zhou*[1,2,6,7,*]

[1] Institute of Fundamental and Frontier Sciences, University of Electronic Science and Technology of China, Chengdu 611731, China

[2] Key Laboratory of Quantum Physics and Photonic Quantum Information, Ministry of Education, University of Electronic Science and Technology of China, Chengdu 611731, China

[3] Yingcai Honors College, University of Electronic Science and Technology of China, Chengdu 611731, China

[4] Southwest Institute of Technical Physics, Chengdu 610041, China

[5] Institute of Systems Engineering, AMS, Beijing 100141, China

[6] Center for Quantum Internet, Tianfu Jiangxi Laboratory, Chengdu 641419, China

[7] CAS Center for Excellence in Quantum Information and Quantum Physics, University of Science and Technology of China, Hefei 230026, China

**Correspondence:** Yun-Ru Fan (yunrufan@uestc.edu.cn), Kai Guo (guokai07203@hotmail.com), Qiang Zhou (zhouqiang@uestc.edu.cn)

**Abstract:** Quantum networks play a pivotal role in quantum information science, which not only provide a secure communication platform for remote access to quantum computers but also serve as the strategic core for achieving large-scale quantum information processing, forming the foundational infrastructure for the future global-scale quantum internet. Quantum teleportation, which enables the transmission of unknown quantum states over long distances by employing quantum entanglement together with classical communication, is essential for the distribution of quantum resources in the construction of the global-scale quantum internet. To realize a global-scale quantum internet, quantum repeater protocols represent one of the most promising approaches for enabling quantum communication between any nodes. This concise review presents representative experimental demonstrations of quantum teleportation for constructing quantum networks across different physical platforms. Along this trajectory, the review discusses current challenges, open issues, and future perspectives toward scalable and practical quantum internet.

## 1. Introduction

Quantum teleportation[1,2] plays a prominent role in quantum information science and is one of the central techniques in constructing quantum networks[3,4,5]. By employing quantum entanglement as a communication resource, quantum teleportation enables the transfer of quantum states between spatially separated observers, thereby providing an efficient and reliable approach for long-distance quantum information transmission[6]. Such a protocol provides emerging perspectives for understanding the complexity and subtleties of quantum theory, and has also allowed great advances in quantum communication[7,8,9], distributed quantum sensing[10,11] and distributed quantum computing[12,13].

Up to now, quantum teleportation has been demonstrated experimentally in laboratories all over the world across a variety of physical platforms, including photonic qubits, optical modes, atomic ensembles, trapped atoms, and solid-state systems, among others[14]. Among these platforms, photonic qubits naturally act as flying qubits for long-distance quantum communication, facilitating interconnects between remote quantum processors, while solid-state qubits serve as stationary qubits, offering long coherence times for quantum memory and local information processing[15]. Not restricted only to discrete-variable[5,15] qubits, the protocol can also be extended to continuous-variable systems[14,16,17], and their emerging hybrid architectures. While these diverse implementations collectively enrich the quantum communication landscape, the present review maintains a dedicated focus on the progress and networking potential of discrete-variable systems.

Quantum teleportation protocol also forms the foundation of more complex protocols in quantum technologies, such as entanglement swapping[18] and quantum repeaters[19]. In addition, quantum gate teleportation[20,21,22], port-based quantum teleportation[23,24,25], unconditional teleportation[26,27], controlled quantum teleportation[28,29] and teleportation of a shared quantum secret[30] are all derived from the basic scheme of quantum teleportation. The important extension is to a quantum teleportation network consisting of multiple quantum nodes, which can transmit quantum information between any two nodes through teleportation. Nowadays, quantum teleportation has progressed from the demonstration of proof-of-principle to the construction of networks for real-world applications. Against this backdrop, quantum networks have become a focal point of global technological competition, and many countries have formulated strategic plans to gain a technological edge.

In this review, a concise overview of progress in quantum teleportation for the purpose of constructing quantum networks is presented. In Section 2, we describe the protocol and properties of quantum teleportation and give its extension to entangled states, i.e., entanglement swapping. The basic experimental architectures for quantum teleportation networks and quantum repeater architectures toward the realization of a quantum internet are discussed in Section 3. Section 4 presents a review of quantum teleportation experiments, in which the transfer of a quantum state has been demonstrated over free-space links and optical fibre channels. Section 5 focuses on experimental advances in the teleportation of entangled states. Section 6 reviews experimental progress on quantum teleportation across multiple physical platforms, including single atoms, trapped ions, atomic ensembles, diamond nitrogen-vacancy (NV) centers, and absorptive quantum memories. Finally, we conclude this review by discussing open issues, challenges, and future perspectives of teleportation within quantum networks in Section 7.

## 2. Principles and Properties of quantum teleportation

The quantum teleportation protocol usually involves three parties, typically called Alice, Bob, and Charlie (see Figure 1(a)), and relies on both quantum and classical channels. In Figure 1(a), Alice has a particle in a certain quantum state and wants Bob, at a distant location, to make an accurate copy of it without transferring the original particle. According to the No-Cloning theorem[31], no measurement that Alice can perform will yield sufficient information for Bob to prepare a perfectly accurate copy. In 1993, Bennett et al.[1] proposed that the long-range correlations between Einstein-Podolsky-Rosen (EPR) pairs of particles could provide a unique solution to this problem.

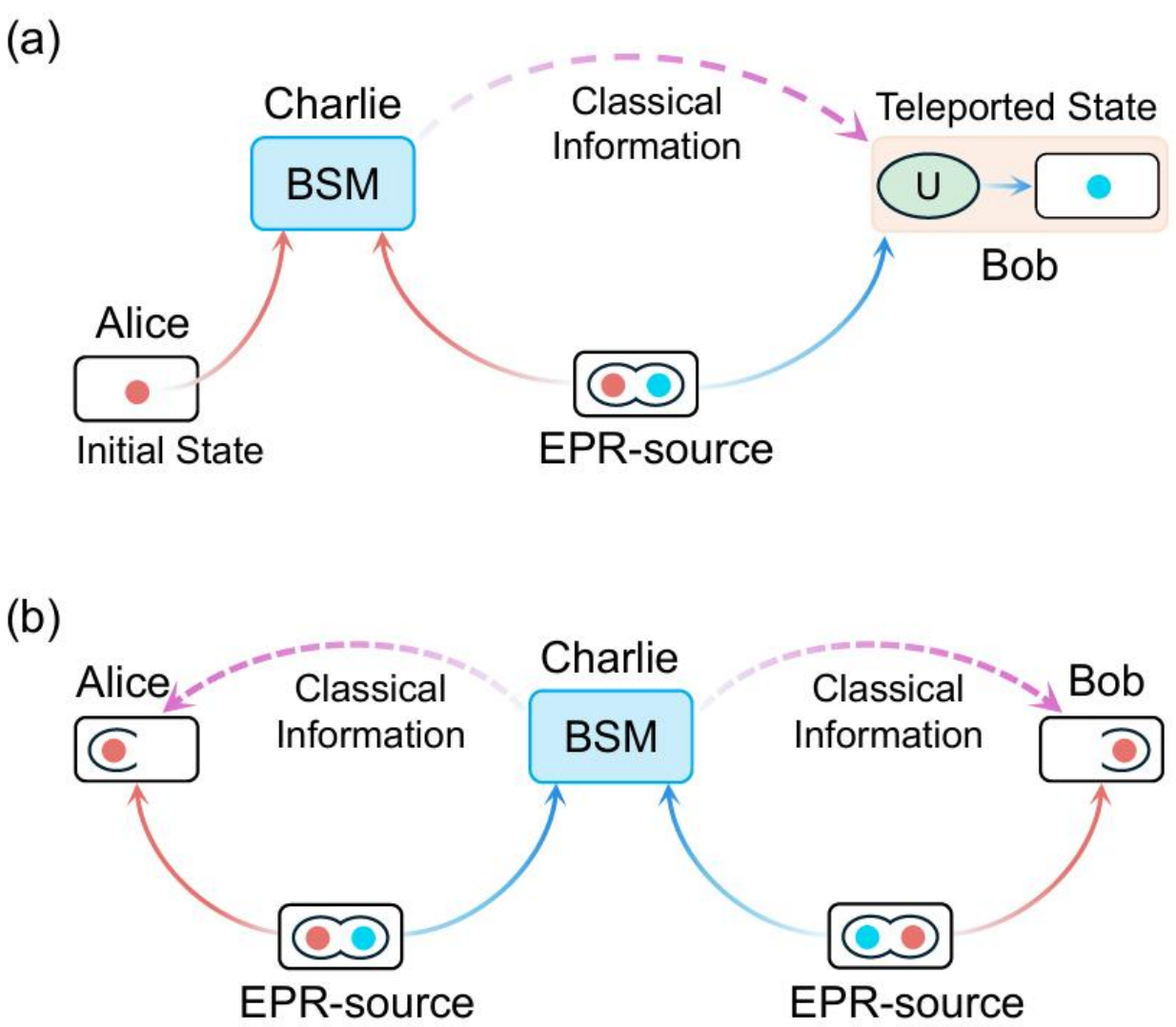

**Figure 1.** Scheme of (a) quantum teleportation of qubit and (b) teleportation of entangled state.

Figure 1(a) shows the scheme of teleportation protocol: suppose particle 1, which is to be teleported by Alice, is in the initial state

$$|\phi\rangle_1 = a|0\rangle_1 + b|1\rangle_1, \tag{1}$$

where $|0\rangle$ and $|1\rangle$ are two orthogonal bases, and a and b are two complex coefficients satisfying $|a|^2 + |b|^2 = 1$. The subscript denotes the corresponding particle. It is obvious that particle 1 here is generally regarded as a coherent superposition of $|0\rangle$ and $|1\rangle$. Two entangled particles 2 and 3 in state

$$|\Psi^-\rangle_{23} = \frac{1}{\sqrt{2}}(|0\rangle_2 \otimes |1\rangle_3 - |1\rangle_2 \otimes |0\rangle_3), \tag{2}$$

are sent to Charlie and Bob, respectively. Here, $\otimes$ denotes the tensor product. Although the violation of Bell's inequalities establishes the possibility of non-classical correlations between Charlie and Bob, the EPR pair at this stage contains no information on the individual particles. Therefore, the whole quantum system can be written as

$$\begin{aligned} |\Phi_{\text{total}}\rangle_{123} &= |\phi\rangle_1 \otimes |\Psi^-\rangle_{23} \\ &= \frac{1}{2}[-|\Psi^-\rangle_{12} \otimes (a|0\rangle_3 + b|1\rangle_3) \\ &\quad -|\Psi^+\rangle_{12} \otimes (a|0\rangle_3 - b|1\rangle_3) \\ &\quad +|\Phi^-\rangle_{12} \otimes (a|1\rangle_3 + b|0\rangle_3) \\ &\quad +|\Phi^+\rangle_{12} \otimes (a|1\rangle_3 - b|0\rangle_3)], \end{aligned} \tag{3}$$

where $|\Phi^\pm\rangle = \frac{1}{\sqrt{2}}(|0\rangle \otimes |0\rangle \pm |1\rangle \otimes |1\rangle)$ and $|\Psi^\pm\rangle = \frac{1}{\sqrt{2}}(|0\rangle \otimes |1\rangle \pm |1\rangle \otimes |0\rangle)$ are the four Bell states.

Then, Charlie performs a Bell state measurement (BSM)[32,33,34,35] between particles 1 and 2 and randomly detects one of the four Bell states with an equal probability of 25%. Quantum physics predicts that once particles 1 and 2 are projected onto Bell states, the state of particle 3 at Bob is completely projected onto the initial state up to a unitary operation. When Charlie informs Bob of the result of the BSM via the classical channel, Bob can perform a result-dependent operation, i.e., one of the unitary rotations $\{I,Z,X,ZX\}$ ($I$ denotes the identity, and $Z$ and $X$ are the Pauli operators), and thus obtain particle 3 in the initial state of particle1.It is important to note that particle 1 on Charlie's side is no longer in its original state due to the BSM. Therefore, particle 3 is not a clone, but the result of real teleportation.

No information about the particles is revealed by BSM, therefore neither Charlie nor Bob obtains any information about the state, leading to the fact that quantum teleportation escapes the verdict of the No-Cloning theorem. Moreover, it is not necessary for Charlie to know where Bob is, he can still teleport the quantum state to him by broadcasting the classical information to all the places he might be. Note that Bob can only complete the teleportation after the arrival of the classical message sent by Charlie, so there is no superluminal communication in quantum teleportation. Owing to the constraints imposed by the No-Cloning theorem, conventional repeater strategies in classical telecommunication systems, such as optical amplification, cannot be directly applied to the long-distance transmission of quantum information. Fortunately, the quantum repeater scheme provides a promising avenue to overcome this limitation by employing teleportation of quantum states and quantum entanglement, which enables the realization of a global-scale quantum internet.

One important application of quantum teleportation is the teleportation of entangled states, also known as entanglement swapping[18], as shown in Figure 1(b). Suppose the first EPR-source generates particles 1 and 2 in the Bell state $|\Psi^-\rangle$, and the second EPR-source generates particles 3 and 4 also in $|\Psi^-\rangle$. Particles 1 and 4 are sent to Alice and Bob, respectively, while particles 2 and 3 are sent to Charlie for BSM. Therefore, the whole quantum system for entanglement swapping can be represented as

$$\begin{aligned} |\Phi_{total}\rangle_{1234} &= |\Psi^-\rangle_{12} \otimes |\Psi^-\rangle_{34} \\ &= \frac{1}{2}[|\Psi^+\rangle_{23} \otimes |\Psi^+\rangle_{14} \\ &\quad - |\Psi^-\rangle_{23} \otimes |\Psi^-\rangle_{14} \\ &\quad - |\Phi^+\rangle_{23} \otimes |\Phi^+\rangle_{14} \\ &\quad + |\Phi^-\rangle_{23} \otimes |\Phi^-\rangle_{14}]. \end{aligned} \tag{4}$$

Quantum physics predicts that once the BSM is performed on particles 2 and 3, particles 1 and 4 will be projected into one of the four Bell states, each with an equal probability of 25%, as given in Equation (4). By broadcasting Charlie's measurement result, Alice and Bob can determine which Bell state their particles (1 and 4) occupy. Furthermore, any of these four Bell states can be converted into another via local unitary operations. The protocol implementation provides a maximal fidelity of the unit in successful cases. In contrast to transmitting a single quantum state from one qubit to another, entanglement swapping could generate entanglement between two independent qubits that have never interacted in the past. Therefore, this protocol represents an integral constituent of quantum repeaters and a key resource for numerous quantum-information applications that have been implemented in many different systems.

## 3. Quantum teleportation network architectures toward the quantum internet

Based on the quantum teleportation protocol, six fundamental experimental architectures for quantum teleportation networks can be constructed (see Figure 2). In general, a complete quantum teleportation process requires several essential components: quantum state preparation, BSM and the corresponding classical communication of its results, generation and distribution of entangled photon pairs (signal photon and idler photon) via an EPR-source, as well as unitary operations conditioned on the BSM outcomes. The principal advantage of quantum teleportation, as opposed to the direct transmission of qubits, lies in its ability to transfer information without physically transmitting the information-bearing particle itself. This process effectively mitigates the distance limitations imposed by transmission loss and decoherence, due to the non-local correlations of quantum entanglement. Nevertheless, the maximum achievable teleportation distance remains constrained by the distribution range of the entangled particle pairs. In addition, achieving efficient and high-rate quantum teleportation further requires high-precision time synchronization[36].

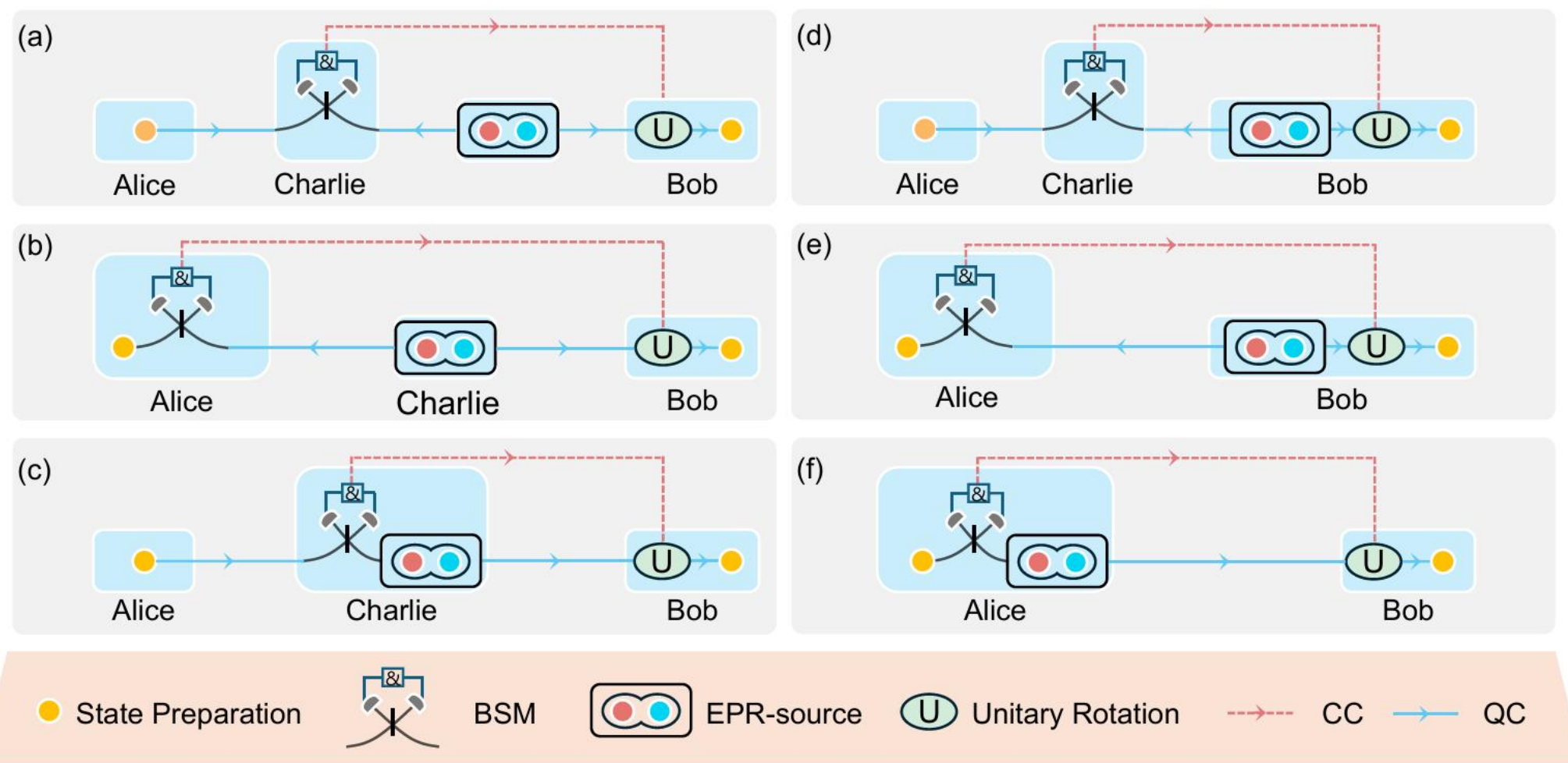

**Figure 2.** Six basic network architectures for quantum teleportation. By distributing the four key functions of the teleportation protocol across different quantum nodes, six network types can be formed. In the quantum network, each fundamental component is interconnected via quantum channel (QC), such as optical fibres or free-space links. The BSM results (classical signals) can be transmitted to the receiver (Bob) via the classical channel (CC).

In the network architecture shown in Figure 2(a)[37,38,39,] Alice prepares the quantum state to be teleported, while the entangled photon pair generated by the EPR-source is distributed to Charlie and Bob, respectively. Typically, the signal photon at Bob's node can only undergo the corresponding unitary rotation after receiving the BSM result from Charlie. Therefore, in the absence of a quantum memory, the transmission distance of the signal photon in the third channel segment must generally exceed that of the idler photon sent to Charlie, ensuring that the signal photon remains available for manipulation once the BSM is completed. To fully exploit the advantage of the non-local property of the entangled state, the spatial separation between Charlie and Bob can be maximized to extend the teleportation distance. However, photon loss still fundamentally limits this distance in the transmission channel. Furthermore, the network architecture can be connected to multiple users via optical switches and utilize Q-chips[40] for routing and addressing, forming a star-topology network[41,42] to enable multi-user quantum information transmission. In the second network architecture shown in Figure 2(b)[43,44,45,46], similar to the first scheme, Charlie (in Figure 2(a)) also acts as a user node for quantum information transmission in a multi-user configuration. In the third experimental architecture shown in Figure 2(c)[41,42,47], the idler photon from the entangled pair prepared by Charlie is sent to the BSM, which then awaits the incoming quantum state from Alice to complete the measurement. At this moment, the signal photon transmitted to Bob collapses into the corresponding quantum state determined by the BSM outcome.

As illustrated in Figure 2(d) and 2(e), the fourth[48,49,50] and fifth[51] experimental architectures employ entangled photon-pair generation at Bob's node, where the idler photon is directed to the BSM module while the signal photon is stored locally in a long optical fibre spool or an alternative quantum memory device. Even if the signal photon is directly operated on or measured before the BSM is completed, without passing through a local fibre delay line, successful teleportation can still be demonstrated. It should be noted that both architectures are operated a posteriori and that probabilistic, effectively instantaneous quantum computations may be enabled, owing to the equal (1/4) probability of all BSM outcomes[52]. Additionally, while fifth architecture has no direct transmission of qubits and the teleportation distance corresponds directly to the spatial distance between Alice and Bob, the transmission loss in a single-link setup is greater than that in a multi-link configuration[50]. In the sixth architecture as illustrated in Figure 2(f)[53], Alice performs the BSM between the prepared quantum state and the idler photon, while the signal photon is sent to Bob. Since the state preparation, BSM, and entanglement generation all occur at Alice's node, the signal photon collapses into the corresponding quantum state upon completion of the BSM. Even though the state transfer distance can be extended, the teleportation distance remains limited to the laboratory scale. This indicates that, although the network architecture employs the quantum teleportation protocol, its performance in terms of transmission loss shows only minor improvement compared with direct qubit transmission. Moreover, single-link network architectures (the fifth and sixth architectures ) exhibit inferior scalability and transmission performance compared with multi-link architectures due to channel attenuation[50].

It follows that the performance of quantum teleportation systems is governed by the constraints of their underlying network architecture. In general, a quantum service provider (Charlie) tends to retain the costly core elements, such as the BSM device, at the network hub, so that the user-side devices can remain as inexpensive as possible. In free-space links, some factors such as atmospheric turbulence, beam diffraction, and weather conditions can significantly increase photon transmission loss. Currently, the construction of short-distance free-space quantum teleportation networks at low altitudes or within urban areas has met the conditions required for practical implementation. Moreover, satellite-based entanglement distribution schemes can effectively extend the teleportation distance and are expected to become the dominant network architecture for realizing global-scale quantum networks in the future, compared with other experimental architectures. By contrast, two-node architectures are more susceptible to photon attenuation than multi-node configurations due to increased transmission loss along the optical fibre. As such, the quantum teleportation networks are mostly still confined to metropolitan areas. To evolve from metropolitan quantum networks to a global quantum network, one approach is to seek low-loss optical channels, such as vacuum beam guides[54] and hollow-core optical fibres[55]. Moreover, the quantum repeater scheme, which can reduce photon loss in optical fibres from exponential to polynomial scaling, is regarded as the most promising approach for building the quantum internet. For real-world deployment, the foundational network architecture can be expanded into duplex, multi-user star-topology networks and hybrid-channel quantum networks[56], ultimately enabling the establishment of a practical quantum internet on a global scale through quantum repeater schemes.

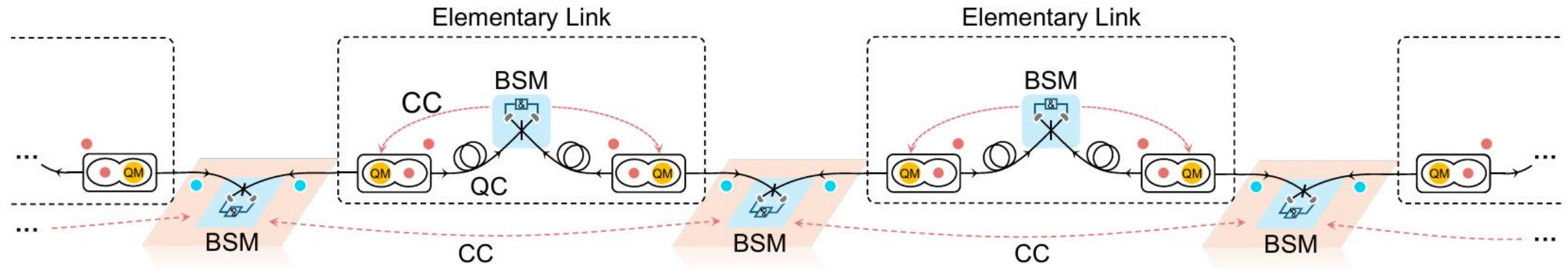

**Figure 3.** Quantum repeater architecture and the elementary link with quantum memories. The long transmission distance can be divided into several elementary links and connected by multiple rounds of entanglement swapping. Each elementary link consists of two nodes located at the ends, where each node can establish entanglement between photons and quantum memories either through built-in-type quantum memories or by employing an independent EPR-source combined with absorptive quantum memories. In each elementary link, a successful BSM heralds that entanglement has been established and stored between the quantum memories of the two quantum nodes. The entangled states held in the quantum memories are retrieved and proceed to a BSM only after heralding signals are received by both memories of two adjacent elementary links. Thereby, entanglement is established between the end nodes across the two links, which can then be extended to other links to ultimately establish entanglement between the end points of the whole network[57,58]. QM: quantum memory.

In a prospective global-scale quantum internet, quantum information can be transferred between any two spatially separated quantum nodes via quantum teleportation protocol. The implementation of long-distance quantum teleportation requires the prior distribution of entangled particles to two distant quantum nodes. For that purpose, quantum repeaters[19,59] are regarded as the most promising approach to extend quantum communication over longer distances through teleportation of entangled states and quantum memory. Furthermore, the entanglement purification[60,61] technique can enhance the fidelity of entangled states degraded by factors such as channel noise and decoherence (not shown in Figure 3). The teleportation of entangled states, also known as entanglement swapping, can establish entanglement between quantum nodes for the elementary link. The initial step necessitates the establishment of entanglement between photon and quantum memory. Based on their operating mechanism, quantum memories can be categorized into two fundamental types: built-in-type quantum memories (also known as emissive quantum memories)[62,63,64,65,66] and absorptive quantum memories[58,62,67,68,69,70]. The former enables the inherent, direct entanglement of the quantum memory (hosted in various physical platforms[62], including cold atomic ensembles and single quantum systems) with an emitted photon. The latter, in contrast, is a separate architecture where the quantum memory and the EPR-source are independent components. It is achieved by storing one photon from an entangled photon pair generated by an independent EPR-source into the absorptive quantum memory. The quantum memory stores the entangled state and waits for further entanglement swapping, thereby avoiding the requirement to establish entanglement simultaneously across the elementary links. Once heralding signals are received at both memories, the stored entangled states are retrieved and entanglement swapping is performed via a BSM at the intermediate station, thereby connecting adjacent elementary links and distributing entanglement over a longer distance (see Figure 3). With the successful prior distribution of entanglement, the quantum state will be teleported to any quantum node within the global quantum internet via the teleportation protocol, based on the non-local nature of entangled states. The technical details of various quantum repeater protocols have already been covered in other reviews[5,19,59,62].

**4. Quantum teleportation of a quantum state**

In the laboratory, the preparation of unknown input quantum states, the generation of entangled states, complete BSMs, and active feed-forward of measurement results are all critical factors for the successful experimental realization of quantum teleportation. For constructing quantum networks, photonic qubits are more suitable as flying qubits for teleportation due to the low transmission loss and robustness against environmental decoherence, while matter qubits serve as local quantum information memory or quantum information processing in quantum nodes. Recent experimental advances in quantum teleportation for building quantum networks will be summarized in the following.

The pioneering experimental demonstration of quantum teleportation was carried out by Zeilinger's group[2] in 1997, providing the first empirical confirmation of its feasibility and laying the groundwork for the future development of a quantum internet on a global scale. In the early stages of research, the demonstration of quantum teleportation using photon qubits[26,71] or matter qubits[72,73] remained at the proof-of-concept stage, with state-transfer distances generally not exceeding 0.1 km in the laboratory. Subsequently, the state-transfer distance of free-space quantum teleportation was progressively extended. Representative demonstrations include a ground-to-ground free-space link of approximately 16 km near Beijing[44], a link exceeding 97 km across Qinghai Lake[45], and a cross-sea link exceeding 143 km between La Palma and Tenerife in the Canary Islands[47] (see [74] for a detailed review). Based on previous exploratory studies[75,76,77,78], satellite-based implementations with the Micius quantum science satellite achieved quantum teleportation from ground to satellite (uplink) over distances of 500-1400 km[53], and quantum-state transfer between two ground stations via the satellite covering a total distance of 1200 km[46]. These experiments successfully demonstrated the transfer of polarization-encoded quantum states between distant nodes via free-space channels, achieving teleportation fidelities that consistently exceed the classical limit of 2/3[79] .

Advances in free-space quantum teleportation are driven by the small photon absorption of the atmosphere within certain wavelength ranges, combined with negligible transmission losses and minimal polarization decoherence in outer space, making free-space links a suitable option for long-distance photon transmission. Moreover, the extremely weak birefringence of the atmosphere ensures a high degree of preservation of polarization entanglement, so polarization-encoded quantum states are commonly used in free-space quantum communication. The integration of quantum satellites with ground stations may enable the realization of a global-scale quantum network via free-space channels.

Quantum teleportation over optical fibre is also a critical step in the roadmap to realizing global-scale quantum networks, in which optical fibres serve as quantum channels, remaining immune to adverse weather conditions while avoiding the use of large-aperture optics and other complex techniques. The transmission loss of photons in standard single-mode fibre has been reduced to approximately 0.2 dB/km at telecommunication wavelengths, making it ideal for long-distance quantum teleportation. Benefiting from the highly developed fibre-optic communication industry, the practical implementation of quantum networks can also take advantage of numerous well-established fibre-optic network resources and proven communication devices and technologies.

The first experimental demonstration of long-distance quantum teleportation over optical fibre was achieved by Gisin's group[37]. Notably, qubits and necessary entangled states are encoded in superposition and entanglement of time-bins[81,82] in this experiment, such encoding is more robust against the decoherence effect in optical fibre. Subsequently, building on their previous experiment, the team extended the total transmission distance to 6.2 km by inserting long optical fibres[38]. Although high-fidelity teleportation of polarization qubits was achieved over an 800 m optical fibre beneath the River Danube[43], time-bin qubits remain the preferred choice for realizing long-distance quantum networks in optical fibre. The longest transmission distance of photons in quantum teleportation over optical fibre has been achieved more than 100 km[82].

The most important extension for quantum teleportation is to a quantum teleportation network, where photons are usually created and processed in remote nodes. It requires independent quantum sources, prior entanglement distribution[39,83], and active feedforward operation, where these elements should be implemented simultaneously in a single experimental realization of quantum teleportation. Two independent quantum teleportation experiments over intercity optical-fibre networks in Hefei[41] and Calgary[51] marked an important milestone toward fibre-based quantum networks (see [15,71] for a detailed review). In 2020, Valivarthi et al.[48] demonstrated quantum teleportation of time-bin qubits at a telecommunication wavelength of 1536.5 nm, achieving an average fidelity of at least 90 % over 44 km of single-mode fibre using advanced fibre-coupled devices. Building on previous work[84], the author's group[49] experimentally demonstrated metropolitan quantum teleportation of time-bin qubits over a 64 km optical fibre channel (see Figure 4(a)). This system employed a high-performance entanglement source and achieved a state-transfer rate of $7.1 \pm 0.4$ Hz together with an average single-photon fidelity exceeding $90.6 \pm 2.6$ %, marking a significant step toward scalable and high-rate metropolitan quantum networks.

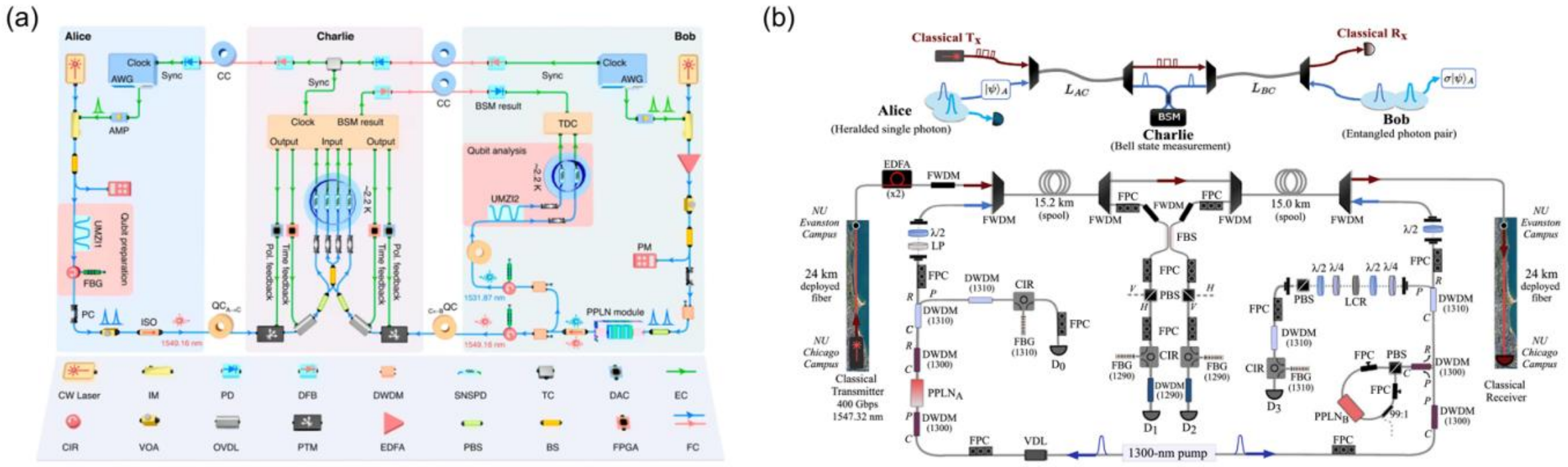

**Figure 4.** (a) Hertz-rate three-node quantum teleportation system over a 64 km optical fibre network[49]. (b) Quantum teleportation experiment coexisting with classical communication in a 30.2 km optical-fibre link[50]. Panel (a) reproduced from ref. [49] and panel (b) reproduced from ref. [50], both under a Creative Commons license CC BY 4.0.

In 2024, Thomas et al.[50] reported an experiment on quantum teleportation coexisting with classical optical communication in the same optical fibre. As shown in Figure 4(b), O-band quantum light (1290 nm) was co-propagated with a 400-Gbps C-band classical communication signal (1547.32 nm) over 30.2 km of fibre via wavelength-division multiplexing. This work successfully demonstrates the feasibility of coexisting quantum teleportation and classical communication within existing fibre infrastructure, while also indicating that leveraging current optical networks for constructing a quantum internet could substantially reduce deployment costs. In 2025, Zhang's group[42] achieved chip-to-chip photonic quantum teleportation of time-bin qubits over 12.3 km of optical fibre, which demonstrates the potential of chip integration for scalable quantum networks. These works underpin the development of quantum repeater-based communications and mark a critical step toward a long-distance quantum network in the real world.

**5. Quantum teleportation of entangled state**

Quantum teleportation of entangled states represents a significant extension of quantum teleportation protocols, where the core concept involves establishing quantum entanglement between particles that have never directly interacted by performing joint measurements on particles originating from different entanglement sources. This process enables the establishment of quantum entanglement between remote nodes and is therefore widely regarded as a core enabling protocol for quantum relays (without quantum memories), quantum repeaters (with quantum memories), and, more broadly, the realization of a quantum internet. The first experimental demonstration of entanglement swapping was reported by Zeilinger's group in 1998[89]. In the subsequent improved experiment, the presence of non-local quantum correlations was unambiguously verified by observing violations of Bell's inequalities[71]. Delayed-choice entanglement swapping provides a particularly instructive perspective on the nature of quantum entanglement[90,91], showing that the presence or absence of entanglement between two remote particles is determined by the measurement performed on their partners, even when this measurement is chosen after the remote particles have been detected[92,93]. This type of delayed-choice gedanken experiment has been discussed in another review[52].

Concurrently, entanglement swapping experiments have progressively evolved from proof-of-principle demonstrations toward long-distance, scalable, and high-fidelity quantum network implementations enabled by advanced techniques and architectures. In 2005, Riedmatten et al.[94] reported entanglement swapping of time-bin entangled photons over more than 2 km of optical fibre using separated entangled sources. Subsequently, multiple experiments have successfully conducted further experimental demonstrations of entanglement swapping by enhancing detection schemes and the performance of entangled light sources, while optimizing key experimental parameters[95,96,97]. To assess the feasibility of quantum relay without quantum memory, Goebel et al. demonstrated two-stage entanglement swapping using polarization-entangled photons[98]. Quantum state tomography performed prior to the swapping process revealed no measurable entanglement between the relevant photons, whereas after the two-stage swapping operation the expectation value of an appropriate entanglement witness became negative. This observation unambiguously confirms that entanglement was generated through the entanglement swapping process itself. Building upon this foundation, Pan's group[99] experimentally demonstrated fault-tolerant two-hierarchy entanglement swapping using linear optics. This work further reduced system noise and advanced the implementation of fault-tolerant quantum repeater schemes. To realize scalable quantum repeaters with quantum memory, Jin et al.[100] developed a memory-compatible entanglement swapping process. The two 795 nm photons that ultimately became entangled could be directly stored in corresponding quantum memories. The teleportation of an entangled state was experimentally demonstrated over a 143-km free-space link between the Canary Islands of La Palma and Tenerife[101].

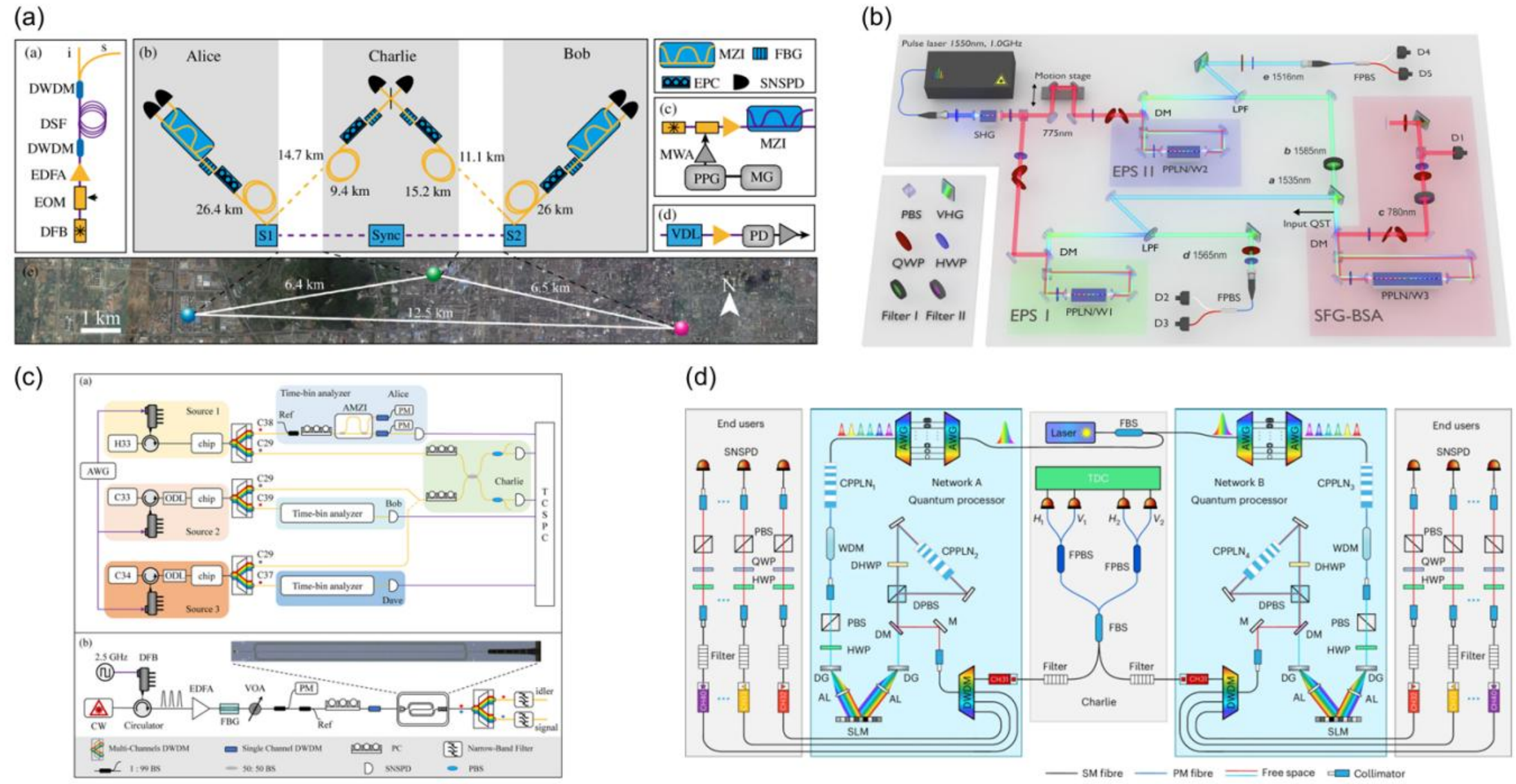

**Figure 5.** (a) Entanglement swapping experiment over 100 km optical fibre with independent time-bin entangled photon-pair sources[85]. (b) Experimental setup for entanglement swapping based on a sum-frequency generation technique[86]. (c) Schematic of cross-channel entanglement swapping between independent on-chip photon sources[87]. (d) Experimental setup for multi-user entanglement swapping-based fusion of two fully connected quantum networks, where independent networks are linked via BSM to establish inter-network entanglement[88]. The panel (a) is reproduced from ref. [85], Copyright © 2017 Optica Publishing Group under the terms of the Open Access Publishing Agreement. The panel (b) is reproduced from ref. [86] and the panel (c) is reproduced from ref. [87] under a Creative Commons license CC BY 4.0. The panel (d) is reproduced with permission from ref. [88], Copyright © 2026, Springer Nature.

To date, entanglement swapping has been successfully demonstrated over optical fibre links exceeding 103 km[85], building upon earlier experimental realizations in a 25.3 km Hefei metropolitan fibre network[102]. As shown in Figure 5(a), the system deployed two independent autonomous 1 GHz sequential time-bin entangled photon-pair sources based on dispersion-shifted fibre and established entanglement between spatially separated photons via a BSM performed at an intermediate third node. Consequently, an interference pattern with a visibility of approximately 73.2% was achieved, which exceeded the bound for demonstrating entanglement. This work validates the feasibility of such technologies for long-distance quantum networks, providing an experimental foundation for their future construction and for fundamental tests of quantum theory.

In photonic entanglement swapping experiments, BSMs are most commonly performed using linear optics. However, this approach only allows for partial Bell-state discrimination and requires post-selection on specific measurement outcomes for practical implementation, even when the efficiency of linear-optics BSMs is enhanced through auxiliary photons. In contrast, performing BSMs via nonlinear interaction processes provides a new route toward high-efficiency, high-fidelity entanglement swapping[103]. Sum-frequency generation enables the nonlinear conversion of photons from two entangled sources into a single higher-frequency photon, heralding successful entanglement swapping. At the same time, this process suppresses multi-photon emissions from the sources, thereby enhancing the fidelity of the swapped entanglement[103]. Combining this technique with dense wavelength-division multiplexing, Cheng's group[104] demonstrated a multi-user entanglement distribution network based on energy-time entanglement. By performing near-ideal entanglement swapping via sum-frequency generation, all four Bell states were unambiguously identified without post-selection, enabling high-fidelity entanglement between any pair of users within the four-user network. In 2025, Tsujimoto et al.[86] reported an entanglement swapping experiment based on the nonlinear process of sum-frequency generation. The core of this work was the development of a high signal-to-noise, stable sum-frequency generation Bell-state analyzer (the experimental setup is shown in Figure 5 (b)). By combining a stable Sagnac interferometer with superconducting nanowire single-photon detectors exhibiting an ultralow dark count rate, this unit met the stringent signal-to-noise requirements for BSMs utilizing single-photon nonlinearity. Notably, this work realized entanglement swapping via sum-frequency generation between two independent single photons, providing a key technological paradigm for faithful, all-photonic quantum information processing. Despite its advantages over linear-optics approaches, the efficiency of the sum-frequency generation process still requires significant improvement to meet practical demands.

The maturity of integrated circuit technology is enabling the development of on-chip integrated quantum light sources, which are key to building scalable, stable, and highly integrated quantum networks. In 2021, Samara et al.[105] generated entangled photon pairs via spontaneous four-wave mixing in integrated silicon nitride microring resonators and demonstrated entanglement swapping between two independent asynchronous sources, achieving a visibility of $91.2 \pm 3.4\%$. In 2026, Wang et al.[87] proposed and realized an on-chip entanglement-swapping scheme utilizing low-loss silicon waveguides and frequency-offset pumps, with the core setup shown in Figure 5(c). The design employed tailored waveguide structures to minimize loss and used frequency-offset pumping to enable noise-suppressed, cross-channel operation. This achieved a high-speed entanglement swapping rate of 207 events per hour while maintaining visibilities above 90%. These work sets a benchmark for chip-based entanglement swapping, providing a viable solution for real-world quantum communication networks.

Toward the full connectivity requirements for independent quantum networks, Huang et al.[88] reported a quantum fusion technique based on multi-user entanglement swapping between two independent quantum networks. As shown in Figure 5(d), through an active temporal and wavelength multiplexing scheme, the entangled state is initially shared between the quantum processor and the user end. Following BSM between the two quantum processors, 18 users from different networks can establish polarization-based quantum entanglement with each other, providing a novel approach for quantum entanglement interconnection across independent quantum networks.

Apart from the above, entanglement swapping also demonstrates the potential for constructing complex quantum networks in multi-photon entangled states[106,107], discrete and continuous variable systems[108,109,110,111,112], and high-dimensional quantum states[113]. Despite the variety of experimental approaches, these experimental advances highlight the potential applications of quantum repeater protocols, which will ultimately lay the foundation for practical long-distance quantum communication networks. Consequently, the transmission of quantum information between arbitrary nodes can be achieved through quantum teleportation within the quantum network with a fully connected logical topology.

## 6. Quantum teleportation with quantum memory

The integration of quantum teleportation with quantum memories constitutes a key enabling element for the realization of scalable quantum repeaters. In particular, quantum teleportation of entangled states between quantum memories at two nodes establishes long-distance entanglement, forming the fundamental links of a scalable quantum repeater network. As the core component of the future global quantum internet, quantum memory enables the long-term storage and on-demand retrieval of quantum information, and has achieved a series of significant experimental breakthroughs in recent years. Currently, quantum repeaters are primarily being developed along two technical paths based on the classification of built-in-type quantum memory and absorptive quantum memory, with multiple implementation platforms emerging in parallel. Owing to their distinct underlying mechanisms, different physical platforms exhibit varied performance across key metrics, including entanglement generation efficiency, storage lifetime, fidelity, transmission distance, and system scalability. This section focuses on the principal physical implementations underpinning quantum repeaters, and reviews experimental progress in quantum teleportation and entanglement swapping with quantum memories across different platforms.

**Single atom.** Single neutral atom is considered a key representative of built-in-type quantum memory, due to the advantages of long coherence times, excellent controllability, and highly efficient photon coupling. In typical schemes, single atoms are confined within high-quality optical cavities or optical dipole traps, where cavity-enhanced light-matter interactions enable a high-fidelity quantum interface between the atomic spin state and single-photon polarization states or time-domain modes, supporting high-fidelity atom-photon entanglement. In 2012, Hofmann et al.[114] reported the first experimental demonstration of heralded entanglement between two $^{87}Rb$ atomic spins separated by 20 m. Each node independently prepared a maximally entangled atom-photon state, and the photons were transmitted to a central station for BSM. Successful detection events heralded the establishment of remote atom-atom entanglement via entanglement swapping. By leveraging cavity-enhanced light-matter interactions, Nölleke et al.[115] achieved high-efficiency state mapping, enabling unconditional quantum teleportation between independent nodes. This demonstration underscored the capability of single-atom interfaces to not only distribute entanglement but also perform elementary quantum information processing tasks.

The scalability of single-atom networks has seen rapid expansion, moving from short-range proof-of-principle tests to long-haul functional links. In 2017, an event-ready Bell test was realized between two rubidium atoms separated by 395 m[116], establishing the framework for heralded entanglement distribution. As shown in Figure 6(a), van Leent et al.[117] in 2022 demonstrated heralded entanglement between two independent single atom quantum memories over a 33 km fibre link. By employing quantum frequency conversion to shift atomic photons to telecom wavelengths for reducing transmission loss and performing a BSM at a central station, entanglement swapping was achieved. Although the fidelity and generation rate remain limited by memory decoherence and low collection efficiency, this work verifies the feasibility of quantum frequency conversion and entanglement swapping over metropolitan scale fibres, thereby laying an experimental foundation for future quantum repeater networks. Most recently, in 2026, Lu et al.[118] pushed this frontier to the 100-km scale, achieving the first device-independent quantum key distribution based on single atoms. By synergizing high-efficiency entanglement swapping, robust phase stabilization, and high-fidelity detection, this milestone demonstrates the transition of single-atom platforms from fundamental entanglement distribution to the practical implementation of complex quantum protocols.

**Trapped ion.** Trapped-ion systems represent a premier platform for quantum network nodes[119], owing to their exceptional coherence properties and high-fidelity gate operations. Typically, individual ions are confined in radio-frequency Paul traps and entangled with single photons through controlled spontaneous emission, establishing robust light-matter interfaces. A seminal milestone was achieved in 2007 by Moehring et al.[64], who demonstrated the first heralded entanglement between two $^{171}Yb^{+}$ ion qubits separated by 1 m. While this work confirmed the feasibility of ion-based remote entanglement, the generation rate remained bottlenecked by low photon collection efficiencies and the probabilistic nature of two-photon interference. Subsequent advancements systematically tackled these challenges and further extended the applicability of the protocol. In 2008, Matsukevich et al.[120] utilized this remote link to demonstrate a violation of Bell's inequality, rigorously verifying that non-local correlations are preserved across spatial separations. This was followed by the first demonstration of quantum state teleportation between distant $Yb^{+}$ ions in 2009 [121], providing a foundational primitive for distributed quantum computing. To enhance entanglement fidelity in non-ideal conditions, Vittorini et al.[122] introduced time-resolved detection and active phase feedforward, effectively mitigating photon frequency distinguishability. Collectively, these breakthroughs culminated in the realization of a modular quantum network architecture[123], establishing trapped ions as a scalable backbone for future quantum internet infrastructures.

In 2020, Stephenson et al. achieved high-fidelity remote entanglement between two $^{88}Sr^{+}$ ions[124], significantly boosting the entanglement rate to 182 $s^{-1}$ with a 94% fidelity through an optimized high-numerical-aperture photon collection geometry. Along with the rate improvements, Langenfeld et al.[125] demonstrated an innovative teleportation protocol in which a single traveling photon sequentially experiences cavity-enhanced reflections at both the transmitter and receiver nodes. This heralded scheme bypasses the requirement for pre-distributed entanglement, simplifying the resource overhead for state transfer. In 2023, Krutyanskiy et al. reported two advances in trapped-ion quantum networks using $^{40}Ca^{+}$ ions. In one[126], they demonstrated heralded entanglement between two ions separated by 230 m via a photonic BSM, achieving a fidelity of 88%. In the other[127], they realized a quantum-repeater node based on two cavity-coupled ions, where an asynchronous protocol that employs ion-qubit memory increased the success probability by a factor of 128 compared to direct transmission, producing a heralded Bell state with 72% fidelity over 50 km of optical fibre. In 2025, Saha et al.[128] pushed the fidelity frontier to $97.0 \pm 0.4\%$ using time-bin photons, while providing a critical analysis of how ion motional coupling imposes fundamental constraints on detection timing windows. As shown in Figure 6(b), Liu et al.[129] in 2026 demonstrated remote entanglement between two ion quantum memories over a 10-km fibre link, achieving a coherence lifetime of 550 ms that exceeds the average entanglement generation time of 450 ms. This result fulfils the event-ready prerequisite for functional quantum repeaters. By leveraging this sustained entanglement, the team further implemented device-independent quantum key distribution, extending the secure communication distance by two orders of magnitude and bridging the gap between laboratory-scale physics and robust network-level applications.

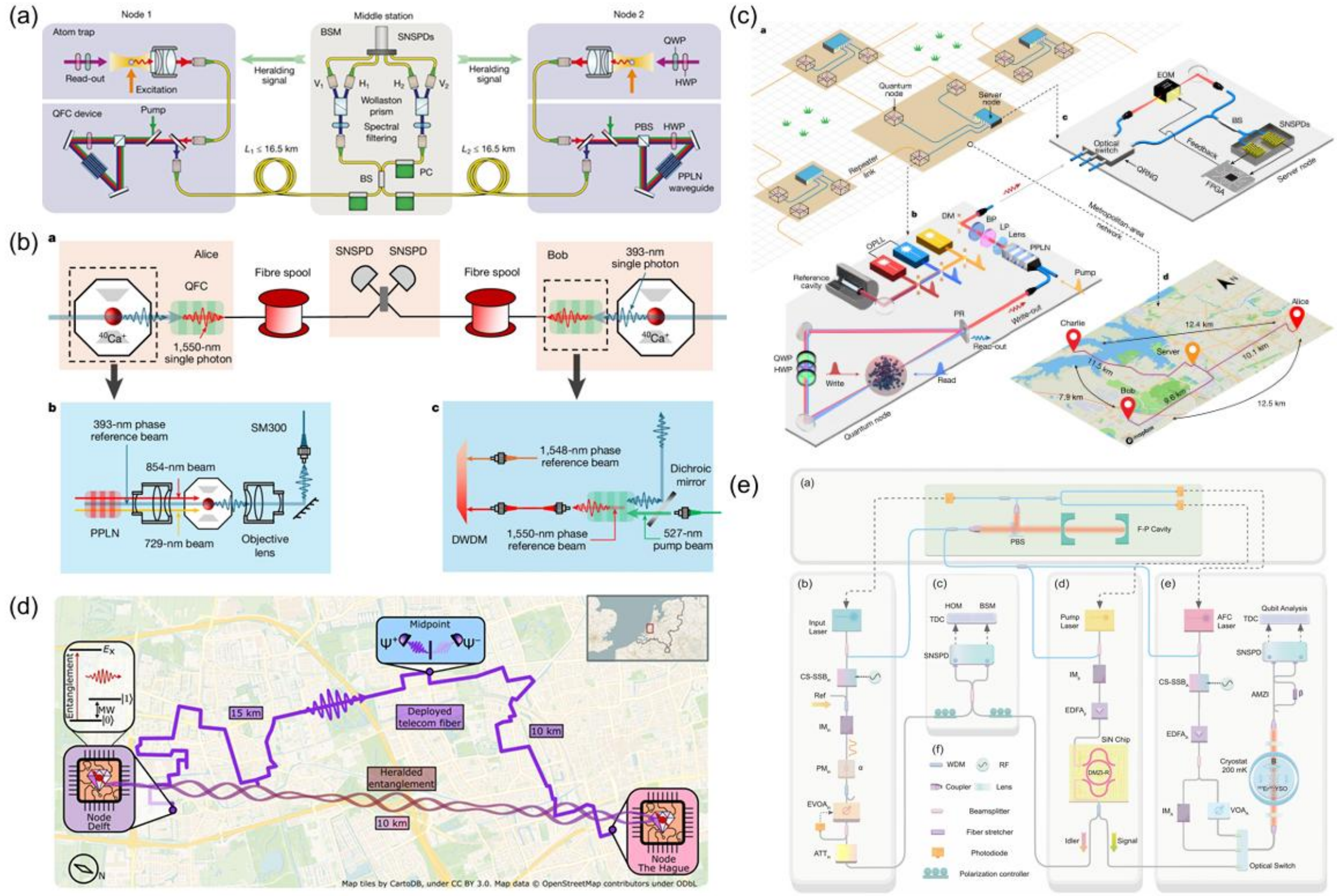

**Figure 6.** (a) Heralded entanglement between single atoms mediated by entanglement swapping over 33 km of telecom-wavelength fibre[117]. (b) Experimental setup for heralded remote ion-ion entanglement in a quantum repeater over 10 km of spooled fibre[129]. (c) A scalable metropolitan quantum network with three nodes, each containing an atomic-ensemble quantum memory, connected via optical fibres of up to 12.5 km to a central server for heralded entanglement generation[130]. (d) Metropolitan-scale heralded entanglement of two independent solid-state qubits (NV centers) separated by 10 km and connected via 25 km of deployed optical fiber to a midpoint station[131]. (e) Quantum teleportation from a telecom-wavelength photonic qubit to a solid-state quantum memory based on erbium-ion ensembles132. The panel (a) is reproduced from ref. [117] under a Creative Commons license CC BY 4.0. Panel (b) is reproduced with permission from ref. [129], Springer Nature Limited. The panel (c) is reproduced with permission from ref. [130], Springer Nature Limited. The panel (d) is reproduced from ref. [131] under a Creative Commons license CC BY 4.0. The panel (e) is reproduced with permission from ref. [132], Copyright (2025) by the American Physical Society.

**Atomic ensembles.** Quantum memories based on atomic ensembles represent a robust alternative and indispensable technological route for scalable quantum networks. Characterized by intrinsic collective enhancement and seamless light-matter coupling, this platform offers favorable scalability for quantum teleportation and entanglement swapping. Unlike single-emitter systems that necessitate high-finesse cavities or high-numerical-aperture optics, atomic ensembles leverage collective spin excitations distributed across a macroscopic atomic population. Under weak-excitation regimes, this mechanism facilitates efficient single-photon emission and heralded storage, supporting the probabilistic yet scalable construction of network nodes. The architectural foundation for ensemble-based repeaters originated from the DLCZ protocol[9]. Its core principle involves generating remote entanglement through the interference of indistinguishable Raman-scattered Stokes photons, whereby a single-photon detection event projects distant ensembles into a non-local entangled state characterized by a delocalized spin-wave excitation. In 2005, Chou et al.[133] reported a seminal implementation, realizing measurement-induced entanglement between spatially separated cold $^{87}Rb$ ensembles. The stored entanglement was subsequently retrieved via an anti-Stokes process, demonstrating a functional light-matter interface. Parallel to this, Sherson et al.[134] achieved quantum teleportation from a propagating light field to a room-temperature cesium vapor ensemble, establishing a high-fidelity mapping of photonic states onto collective atomic excitations. Building on these primitives, Chou et al.[63] realized functional quantum nodes by implementing asynchronous control over dual ensemble pairs, effectively distributing near-maximally entangled states between remote locations as evidenced by the violation of Bell's inequality. This work marked the first integration of entanglement distribution units into a scalable architecture. In 2008, Chen et al.[135] demonstrated the first memory-integrated quantum teleportation between a flying photonic qubit and a stationary atomic ensemble. By mapping an unknown single-photon polarization state onto a collective spin excitation over a 7-meter link, they confirmed that the quantum information could be reliably preserved and retrieved for up to 8 $\mu$s. Concurrently, Yuan et al.[65] demonstrated a complete BDCZ quantum repeater node, successfully merging entanglement generation, storage, and swapping. This milestone transitioned the DLCZ-type protocols from fundamental physics verification to a comprehensive and functional node architecture. Following these advancements, teleportation between distant ensemble memories was first realized in 2012[136], while the another demonstration of deterministic teleportation further enhanced operational efficiency, bridging the gap toward high-throughput quantum communication.

Afterward, quantum networks based on atomic ensembles made significant strides in scalability and transmission distance. In 2020, Yu et al.[66] entangled two $^{87}Rb$ atomic-ensemble memories over a fibre link of tens of kilometers, using cavity-enhanced atom-photon entanglement and quantum frequency conversion to telecom band-validating key capabilities for quantum repeaters. In 2021, Li et al.[137] realized heralded remote entanglement via entanglement swapping in a room-temperature atomic ensemble, removing the requirement for laser-cooled systems while preserving the essential functionality for quantum repeater protocols. As shown in Figure 6(c), Liu et al.[130] in 2024 built a three-node star-topology metropolitan quantum network in Hefei, achieving heralded entanglement between quantum memories up to 12.5 km apart. The quantum memory coherence time exceeded 560 $\mu$s, far longer than the photon round-trip time. The network employed a weak-field phase stabilization method to enable concurrent entanglement generation between any two nodes, and its star-topology architecture can be directly extended to more quantum nodes. In 2025, Wang et al.[138] significantly improved the entanglement swapping efficiency in atomic ensemble systems via spatial multimode multiplexing, demonstrating the potential for scaling to multi-user quantum networks. Additionally, an experimental realization of entanglement preparation using Rydberg superatoms was reported[139], showcasing a new pathway for high-fidelity entanglement generation in strongly interacting systems.

**Solid-state diamond.** The NV center in diamond serves as a versatile hybrid-spin architecture integrating a coherent optical interface with a multi-qubit register within a robust solid-state lattice. By employing dynamical decoupling sequences, NV electron spins can maintain coherence on millisecond scales. At cryogenic temperatures, spin-selective optical transitions facilitate the construction of high-fidelity spin-photon interfaces. In this framework, the electron spin functions as a high-speed communication qubit, while proximal nuclear spins, such as $^{14}N$ and $^{13}C$, provide long-lived quantum memory resources through hyperfine interactions. This hierarchical interface-storage configuration enables the parallel execution of remote entanglement distribution and local state processing. Significant breakthroughs in establishing remote connectivity began in 2013 when Bernien et al.[140] demonstrated the first heralded entanglement between two NV centers separated by 3 m. This experiment employed a measurement-induced protocol analogous to the DLCZ scheme. Subsequently, Pfaff et al.[27] realized unconditional quantum teleportation over 3 m. In this work, a quantum state initially encoded in a $^{14}N$ nuclear spin was reconstructed at a remote node through real-time feedforward control. A historical milestone was reached in 2015 by Hensen et al.[141] who performed a loophole-free Bell test over a 1.3-km link. This achievement simultaneously closed the detection and locality loopholes for the first time. To address the rate bottlenecks inherent in two-photon protocols, Humphreys et al.[142] implemented a single-photon heralding scheme to boost remote entanglement delivery to 39 Hz. This represented a three-order-of-magnitude enhancement over prior benchmarks. Complementing these electronic-link advancements, Tsurumoto et al.[143] demonstrated the teleportation of photonic polarization states directly into $^{13}C$ nuclear spin registers by utilizing the electron spin as a coherent intermediary. Concurrently, unconventional platforms including bulk diamond phonons have been explored for room-temperature state mapping[144]. Collectively, these advancements in integrated solid-state registers demonstrate a viable pathway for high-throughput quantum repeaters and the realization of heterogeneous quantum networks.

The transition from point-to-point links to complex network topologies was catalyzed by the first realization of a three-node solid-state quantum network in 2021[145]. This architecture utilized NV electron spins across three nodes as communication qubits, while the central node incorporated a $^{13}C$ nuclear spin for quantum storage. By integrating phase-stabilized fibre links with real-time feedback, the network successfully demonstrated GHZ state generation and entanglement swapping, providing an essential experimental testbed for the quantum internet control stack. Building on this infrastructure, Hermans et al.[146] achieved quantum teleportation between non-adjacent nodes. Without a direct physical link, the quantum state was teleported from the source to a distant target via middle-node entanglement swapping and active feedforward, marking a shift toward arbitrary-node connectivity. Scaling these networks to metropolitan scales required overcoming the intrinsic attenuation of optical fibres over extended reaches. In 2024, Stolk et al.[131] demonstrated a real-world metropolitan-scale connection of NV centers over a geographical distance of 10 km, using a 25-km link of deployed optical fibre that runs through both fibre spools and underground cables (as shown in Figure 6(d)). In this practical setup, two diamond spin nodes were linked to a central midpoint station, with the fibre spools and buried cables faithfully replicating the conditions of an urban quantum network. The key enabling technology was quantum frequency conversion, which shifted the native NV photons to the telecom L-band, thereby drastically reducing transmission losses and effectively bringing solid-state quantum networks into a real-city environment. Recent efforts have also focused on enhancing the robustness and functionality of the light-matter interface. In 2025, Ito et al.[147] demonstrated a teleportation-based state transfer scheme that exhibited superior resilience against spectral and temporal misalignments compared to conventional interference methods. This progress in noise-resilient interfacing is critical for maintaining stable connections in real-world environments. Furthermore, Reyes et al.[148] demonstrated a complete quantum repeater node function by leveraging electron-nuclear entanglement. In this protocol, a photonic polarization state was absorbed and subsequently re-emitted via local BSMs, realizing photon-to-photon teleportation and establishing the NV center as a functional repeater unit for future long-distance quantum communication.

The experimental progress across diverse physical systems confirms that quantum teleportation and entanglement swapping have been systematically validated on multiple integrated quantum memory platforms. These include high-fidelity light-matter interfaces in single neutral atom and trapped-ion systems; scalable quantum nodes leveraging collective spin-wave excitations in atomic ensembles; and hybrid electron-nuclear spin registers in diamond NV centers. Each platform offers unique trade-offs regarding entanglement generation rates, storage lifetimes, interface bandwidths, and overall network scalability, together forming the foundational technological roadmap for contemporary quantum repeater research. In parallel, the emergence of novel integrated platforms is actively expanding the physical boundaries of quantum network nodes. Nanophotonic solid-state memories have recently demonstrated remote entanglement over telecom-band networks[149], showcasing high compatibility with existing fibre-optic infrastructures. Simultaneously, optomechanical systems utilize mechanical-mode-mediated state manipulation to create quantum interfaces[150], introducing degrees of freedom distinct from conventional atomic or spin-based systems. These emerging platforms not only retain the fundamental ability to store and retrieve quantum states but also provide new avenues for realizing heterogeneous quantum networks, enhancing system integration, and extending the operational reach of quantum repeater architectures.

**Absorptive quantum memory.** In addition to emission-based platforms, absorptive quantum memories represent a distinct and vital technological route for quantum repeaters[151]. The defining feature of these systems is the use of controlled absorption protocols, most notably the atomic frequency comb (AFC), to map and store the quantum state of flying photon[152,153]. Typically implemented in rare-earth-ion-doped crystals[154] or solid-state color centers, absorptive memories offer inherent advantages in terms of broad operational bandwidth, multimode capacity, and native telecom-band compatibility. When integrated with independent entangled photon sources, these memories exhibit significant potential for long-distance quantum networks. A foundational breakthrough occurred in 2014, when Bussières et al.[155] demonstrated quantum teleportation from a telecom-band photon to a solid-state memory. By generating polarization-entangled pairs via spontaneous parametric down-conversion, they stored a signal photon in a crystal while transmitting the idler photon through 25 km of standard fibre, achieving a fidelity of $81 \pm 4\%$. This work established the feasibility of interfacing solid-state absorptive nodes with fibre-optic infrastructures. In 2021, two independent studies achieved a major milestone by demonstrating heralded entanglement distribution using absorptive solid-state quantum memories. Lago-Rivera et al.[156] implemented a heralded entanglement scheme based on cavity-enhanced spontaneous parametric down-conversion sources. In their setup, each node contained a praseodymium-doped ($Pr^{3+}: Y_2SiO_5$) crystal, where the signal photon of a generated pair was stored locally as an AFC excitation, while the corresponding telecom-band idler photon was transmitted to a central station for detection via a fibre beamsplitter. The detection of an idler photon heralded the creation of entanglement between the two remote memories. This scheme further demonstrated temporal multimode operation across 62 modes, highlighting the scalability of the AFC-based approach. Simultaneously, Liu et al.[58] realized remote entanglement using neodymium-doped ($Nd^{3+}: YVO_4$) memories through a two-photon entanglement swapping protocol. In this configuration, idler photons from independent nodes were interfered at an intermediate station performing a BSM, which projected the two distant memories into an entangled state. This approach exhibited strong resilience to fiber phase fluctuations, making it particularly suitable for practical deployment over existing fibre networks. The field of absorptive quantum memories has recently seen rapid advancements in operational rates and cross-platform integration.

In 2023, Lago-Rivera et al.[69] demonstrated active-feedforward-assisted teleportation from a telecom photon to a solid-state qubit over a 1-km fibre link. By leveraging the temporal multimode capacity of the $Pr^{3+}: Y_2SiO_5$ AFC memory, the teleportation repetition rate was increased threefold compared to single-mode schemes, effectively overcoming the constraints imposed by communication latency. In 2024, Iuliano et al.[157] addressed the challenge of heterogeneous node interconnection by developing a two-stage, low-noise quantum frequency conversion system. This allowed the translation of 795-nm photons, which are compatible with thulium or rubidium memories, to the 637-nm emission wavelength of an NV center. They successfully teleported a time-bin encoded state to an NV electron spin with a fidelity of 75.5%, providing a critical interface between disparate physical implementations. In 2025, An et al.[132] demonstrated quantum teleportation from a telecom-wavelength photonic qubit to a solid-state quantum memory based on erbium-ion ensembles, as shown in Figure 6(e). Using a chip-scale silicon nitride microring resonator to generate narrowband entangled photon pairs, they performed a BSM on the idler and input photons, while the signal photon was stored in a $^{167}Er^{3+}:Y_2SiO_5$ crystal whose optical transition lies in the telecom C-band. Quantum state and process tomography yielded an average state fidelity of 81.8% and a process fidelity of 73.6%, both exceeding the classical bounds by more than seven standard deviations. This demonstration highlights the viability of erbium-based quantum memories operating at telecom wavelengths for future long-distance quantum communication and quantum repeater networks. In 2026, Zhu et al.[158] reported the first metropolitan-scale quantum repeater to certify Bell nonlocality, achieving heralded entanglement between two solid-state quantum memories separated by 14.5 km using a two-photon interference scheme combined with temporal multiplexing (1205 modes). The generated Bell state has a fidelity of 78.6 $\pm$ 2.0%, and the CHSH-Bell inequality is violated with S = 2.22 $\pm$ 0.06 (3.7 standard deviations), marking the first certification of Bell nonlocality in a metropolitan quantum repeater. The entanglement distribution rate reaches 0.94 Hz, two orders of magnitude higher than previous single-photon interference demonstrations, while eliminating the need for active phase stabilization of fiber channels, thus providing a practical framework for scalable quantum networks.

In summary, absorptive quantum memories facilitate the efficient conversion from flying photons to stationary qubits by directly mapping quantum states into solid-state media. This platform offers distinctive advantages in telecom-band operation, multimode multiplexing, and long-haul compatibility. When integrated with quantum teleportation and entanglement swapping protocols, these memories are transitioning from individual principal demonstrations toward scalable architectures for practical quantum networks. Collectively, they complement emission-based platforms and drive the evolution of quantum repeater technologies toward the realization of a large-scale global quantum internet.

## 7. Discussion and outlook

Quantum teleportation for constructing a quantum network constitutes a promising, rapidly developing research topic that builds on decades of research into communication technologies and quantum physics. This review introduces the concept and property of quantum teleportation, which is based on the well-known concept of quantum entanglement, called "spooky action at a distance" by Einstein in his EPR paper. Then, it is discussed the foundational architecture of quantum teleportation networks, as well as quantum repeater architecture for constructing large-scale quantum internet. And, the review provides an overview of major experimental advances in quantum teleportation of quantum state and entangled state. Within the framework of quantum repeater architectures, this review systematically synthesizes the major advances toward the realization of a quantum internet, emphasizing the diverse implementation pathways across multiple physical platforms, where quantum teleportation underpins long-distance entanglement distribution and its scalable extension.

Recent results demonstrate that quantum states can be teleported with high fidelity over long distances via both free-space links and optical fibre channels. While the achieved fidelities are sufficient for several practical applications, the current teleportation rates remain limited, primarily due to finite photon collection. Higher repetition-rate photon sources[159], together with continued improvements in coupling and detection efficiencies[160,161], as well as enhanced BSM efficiencies[162,163,164], are expected to significantly increase the overall teleportation event rate. In addition, advanced strategies such as hyperentanglement[165,166,167,168], multiplexing techniques[57,169,170,171,172], and high-dimensional quantum state encoding[113,167,173,174] provide further viable pathways toward enhancing system performance. Additionally, emerging quantum photonic devices, including high-brightness entangled photon sources[175,176], improved BSM modules[164], and low-loss integrated photonic components[87], are expected to further advance the performance and scalability of quantum networks.

Despite successful demonstrations of entanglement swapping and quantum teleportation across diverse physical platforms, the realization of large-scale, cascadable quantum repeater networks remains confronted by significant technical challenges, including the development of large-bandwidth, long-lifetime, and high-fidelity quantum memories, the synchronization of repeater links, and the spectral compatibility of hybrid quantum systems. Several types of solid-state emitters[177,178,179] have been extensively explored as single-photon sources. Hybrid approaches combining free-space and fiber-optic links constitute a critical pathway toward global-scale quantum communication[56,180]. Deploying hybrid classical-quantum communication networks over low-loss optical fibers, such as emerging hollow-core fibers, is poised to establish a new paradigm for large-scale quantum networking[181]. Novel optical topologies[182,183] may offer alternative avenues for advancing quantum networks. Airborne quantum nodes based on drones[184] and high-altitude balloons[185] could provide a practical route to ubiquitous, all-weather quantum network coverage. Toward a genuinely heterogeneous hybrid quantum network, multiband quantum state transfer across microwave and optical frequencies[186] will spectrally interconnect disparate physical platforms, including superconducting qubits, solid-state quantum memories, and atomic systems. Ultimately, all hardware components are expected to be integrated onto photonic chips[42,105,187] and orchestrated by a quantum network operating system[188]. When achieved, such progress will be invaluable for the future quantum internet.

**Author Contributions**

Y. M., Y. F., R. S., Y. Z., Y. L., S. S., Z. Z., Y. W., K. G., G. G., and Q. Z. contributed to the writing and revision of the manuscript. Y. F., Q. Z., and K. G. supervised the work.

**Acknowledgements**

This work was supported by the Sichuan Science and Technology Program (China) under Grant No. 2024YFHZ0370, 2024YFHZ0368, and 2024YFHZ0369; the National Natural Science Foundation of China under Grant No. 62475039, 62405046 and 62375043; the Quantum Science and Technology-National Science and Technology Major Project (China) under Grant No. 2024ZD0300800 and 2021ZD0300701; and the Tianfu Jiangxi Laboratory under Grant No. TFJX-ZD-2025-005.

Received: ((will be filled in by the editorial staff))

Revised: ((will be filled in by the editorial staff))

Published online: ((will be filled in by the editorial staff))